\documentstyle[aps]{revtex}

\title{Quantum fields in disequilibrium: neutral scalar bosons with long-range,
inhomogeneous perturbations}
\author{Mark Burgess}
\date{15 August 1995}
\address{Faculty of Engineering, Oslo College, 0254 Oslo,
Norway\\and\\Institute of Physics, University of Oslo, P.O.Box 1048 Blindern,
0316 Oslo, Norway}
\date{\today}

\begin{document}
\bibliographystyle{unsrt}

\maketitle

\begin{abstract}
Using Schwinger's quantum action principle, dispersion relations are
obtained for neutral scalar mesons interacting with bi-local
sources. These relations are used as the basis of a method for
representing the effect of interactions in the Gaussian approximation
to field theory, and it is argued that a marked inhomogeneity, in
space-time dependence of the sources, forces a discrete spectrum on
the field.  The development of such a system is characterized by
features commonly associated with chaos and self-organization
(localization by domain or cell formation).  The Green functions play
the role of an iterative map in phase space.  Stable systems reside at
the fixed points of the map. The present work can be applied to
self-interacting theories by choosing suitable properties for the
sources. Rapid transport leads to a second order phase transition and
anomalous dispersion. Finally, it is shown that there is a compact
representation of the non-equilibrium dynamics in terms of generalized
chemical potentials, or equivalently as a pseudo-gauge theory, with an
imaginary charge. This analogy shows, more clearly, how dissipation
and entropy production are related to the source picture and transform
a flip-flop like behaviour between two reservoirs into the Landau
problem in a constant `magnetic field'.  A summary of conventions and
formalism is provided as a basis for future work.
\end{abstract}

\section{Introduction}

An increasing number of actual problems in physics find their natural
expression not in the the static (equilibrium) aspects of quantum
systems, but in the kinematical (non-equilibrium) development of their
average properties. Examples include studies of early universe
expansion\cite{calzetta1,pi1,lawrie1}, heavy ion collisions and the
postulated quark-gluon plasma\cite{heinz1}, lasers and other driven
systems\cite{korenman1,senitzky1,senitzky2}, and particle creation in changing
fields of
force\cite{brown1,mottola1}.

The migration from equilibrium to non-equilibrium involves a shift of
paradigm. In common with zero-temperature field theory, particle
systems at equilibrium are often treated by a scattering formalism,
with an initial (in) state and a final (out) state; this is only
sensible if both are known and are at equilibrium with the same
thermodynamic reservoir. The physics of a non-equilibrium system
demands different boundary conditions. The initial and final states
are (by definition) not characterized by the same ensemble and it is
more appropriate to define the state (spectral profile or density
matrix) of the system at some initial time $\overline t_{i}$ and
compute the final state and its consequence at a later time $\overline
t$.  This describes to an initial value problem which is deftly
handled by Schwinger's closed time path (CTP) action principle. The
new picture also implies a concern with probabilities, or expectation
values rather than amplitudes.

In equilibrium, one is used to the notion of translational invariance
in space and time, implying that physical
quantities only depend on the differences of coordinates $x-x'$.  When
the field is driven into disequilibrium, it acquires an additional
dependence on the average position and time $\overline x =
\frac{1}{2}(x+x')$.  This is measured relative to an initial point of
reference $\overline x_{i}$.  In practical applications it is usually
necessary to assume that the dependence on the average coordinate is
{\em quasi-static} or of low adiabatic order in order to make
computations tractable.  The dependence on the average coordinate has important
features: the preservation of unitarity demands that the statistical
state of the field only depend on $\overline x$ and not $x-x'$. Since
the state of the field can only be altered by the intervention of
sources or sinks (hereafter referred to collectively as sources), the
sources must also develop with respect to the average coordinate. Since
one is interested both in fluctuations and the average kinematics, it
is convenient to work with variables and sources which are
bi-local objects rather than working with
the field itself. This is in contrast to the pure field approach used
by Schwinger\cite{schwinger2}. Self-interacting theories are a special case in
which
the field is its own source; they pose mainly calculational
problems---conceptually no new issues are introduced other than
self-consistency.

Since the external sources affect the average state, they can be
regarded as thermodynamical reservoirs, with the caveat that they must
suffer a `recoil' or back-reaction as a result of their effect on the
system. This is not negligeable off equilibrium. Many quantum systems
(the laser, for example) can be treated as two-reservoir systems in
which the `external' reservoir is of comparable magnitude to the local
one.

The nomenclature `open system' is used to describe a system coupled
to independent sources. The name `closed system' is given to
a system without sources or one in which the field is its
own source; in the latter case, the source must become
effectively impotent with regard to the further development
of the average state, as equilibrium is approached.
Equilibrium is only achieved when, either the whole system
approaches some driven limit cycle, or the contact with
external sources is effectively terminated.
In a closed system, the final equilibrium is a thermal state,
or a state of maximum entropy. Some authors define equilibrium
to mean a thermal state rather than merely a static one---and
non-equilibrium to mean anything else. This is somewhat
misleading since a non-thermal but static state is still final
in the absence of new perturbations and must therefore be
considered a point of equilibrium for the system.

The purpose of the present paper is to extend Schwinger's method of
analysis to treat non-equilibrium ensembles of bosons.  This
touches on and extends a number of apparently different approaches to
non-equilibrium\cite{calzetta1,lawrie1,eboli1}. Since Schwinger's
original work \cite{schwinger2} on the initial-value problem, most
authors have been seduced by the functional integral and have
therefore missed the often subtle advantages of Schwinger's
methodology. It is intended that the present work should convey a
pedagogical flavour of the suggested approach, which overlaps with the
existing literature in strategic places without actually following any
of them.  In particular, conventions and definitions (which differ
from most other accounts) have been chosen rather carefully for
practical purposes. Some well known results are rederived in order to
make the present work as self-contained as possible.  The paper begins
with a summarial discussion of the formalism, paying special attention
to the action principle and unitarity; later the most general
quadratic theory which maintains unitarity is presented and the Green
functions are calculated for prescribed sources. Particular attention
is paid to the effects of non-locality in the sources---an issue which
has been largely neglected in previous work, and turns out to place
strong requirements on the behaviour of stable systems. Finally a
brief comparison is made between the present work and other approaches.

\section{Formalism and conventions}

The conceptual framework for the decription of non-equilibrium processes
will include operator field theory, the method of sources
and the local momentum space Green functions. In addition it
proves convenient to use Schwinger's quantum action principle.
This is a statement about the unitary
development of the field with respect to the variation of certain
variables. Since it embodies the equations of motion and the
fundamental commutation relations
for the field, it is both compact and elegant. One begins
with the action operator, which is defined to be the classical
action with the classical field replaced by the field operator, together
with a suitable ordering prescription for the fields.
Here the ordering will be the usual time-ordering and the action
that for a real scalar field without self-interactions.
The Minkowski metric-signature is $(- + + +\cdots)$ which allows
straightforward comparison with the Euclidean theory.

\begin{equation}
S = \int dV_x \lbrace \frac{1}{2}(\partial^\mu\phi)(\partial_\mu\phi) +
\frac{1}{2}m^2\phi^2 - J\phi\rbrace\label{eq:1}
\end{equation}
where $dV_x$ is the Minkowski volume element. The operator
equations of motion now follow from the quantum action
principle\cite{schwinger1}
\begin{equation}
\delta \langle \phi| \phi'\rangle = i \langle \phi|\delta S |\phi'\rangle
\label{eq:2}
\end{equation}
giving
\begin{equation}
\left( -{\vcenter{\vbox{\hrule height.4pt\hbox{\vrule width.4pt
height8pt\kern8pt\vrule width.4pt}\hrule height.4pt}}} + m^2 \right) \phi(x) =
J(x)\label{eq:3}
\end{equation}
Given that $\phi(x)=\phi(x_i)$ at initial time $t_i$ (or, more
generally, on the the space-like hypersurface $\sigma_i$), the solution
to (\ref{eq:3}) may be written
\begin{equation}
\phi(x) = \phi(x_i) + \int_{\sigma_i}^{\sigma} dV_{x'} G_c(x,x')J(x')
\label{eq:4}
\end{equation}
where $\sigma_i$ and $\sigma$ are the initial and final hypersurfaces
and $G_c(x,x')$ is a Green function which satisfies
\begin{equation}
(-\stackrel{x}{{\vcenter{\vbox{\hrule height.4pt\hbox{\vrule width.4pt
height8pt\kern8pt\vrule width.4pt}\hrule height.4pt}}}} +m^2) G_c(x,x') =
\delta(x,x')
\label{eq:5}.
\end{equation}
Both the Feynman propagator and the retarded Green function have this
property.

The surface integral under the variation of the action vanishes
independently implying that the generator of infinitesimal
unitary transformations on the field is\cite{schwinger1}
\begin{equation}
\chi_\phi = \int d\sigma^\mu \phi\partial_\mu\phi.\label{eq:6}
\end{equation}
Since it is easily established\cite{schwinger1} that
the unitary variation of any operator $A$ is
\begin{equation}
\delta A = -i \lbrack A,\chi_A \rbrack\label{eq:7}
\end{equation}
it follows that, on any spacelike hypersurface with
orthogonal vector $\hat{n}^\mu$, one has
\begin{equation}
\lbrack \phi, \Pi_\mu \rbrack \hat{n}^\mu = i\delta({\bf x},{\bf
x'})\label{eq:8}
\end{equation}
with $\Pi_\mu = \partial_\mu \phi$.
This is the covariant statement of the canonical commutation
relations for the field and its conjugate momentum.
To avoid unnecessary notation it is convenient to write this simply as
\begin{equation}
\lbrack\phi({\bf x},t), \partial_t\phi({\bf x'},t) \rbrack = i\delta({\bf
x},{\bf x'})\label{eq:9}
\end{equation}
with the understanding that general covariance is easily restored
by introducing a suitable time-like vector.

{}From the action principle (\ref{eq:2}) it can be shown by
repeated functional differentiation with respect to the
source that

\begin{equation}
\frac{\delta^n\langle\phi_2|\phi_1\rangle_J}{\delta J(x_1)\ldots \delta J(x_n)}
= i^n \langle \phi_2|T(\phi(x_1)\ldots\phi(x_n)) |\phi_1 \rangle\label{eq:10},
\end{equation}
thus the Taylor expansion of the amplitude may be written in the shorthand
form
\begin{equation}
\langle\phi_2|\phi_1\rangle_J = \langle \phi_2|T
e^{iJ\phi}|\phi_1\rangle\label{eq:11}
\end{equation}
where $T$ denotes time-ordering (latest time to the left). This formula
may be regarded as a generating functional for the $n$-point functions
of the theory.
The complex conjugate of this quantity is
\begin{equation}
\langle\phi_2|\phi_1\rangle_J^\dagger= \langle\phi_1|\phi_2\rangle_J = \langle
\phi_1|T^\dagger e^{-iJ\phi}|\phi_2\rangle\label{eq:12}
\end{equation}
where $T^\dagger$ stands for anti-time-ordering (latest field to the right).
This reverse-ordering is necessary to ensure the cancellation of intermediate
fields in the identity:
\begin{equation}
\langle\phi_2|\phi_1\rangle_J \times\langle\phi_1|\phi_2\rangle_J = \langle
\phi_2|T^\dagger e^{-iJ\phi}T e^{iJ\phi}|\phi_2\rangle = 1\label{eq:13}.
\end{equation}
This is the key observation for the construction of the expectation values.
Notice how the operator ordering in (\ref{eq:13}) starts from an early
time, increases to a final time (at the centre of the operator product) and
then reverses back to the initial time. Each field, at each instant along the
closed time path has a mirror counterpart required for the cancellation of the
intermediate operators in (\ref{eq:13}). This property can now be used to
advantage to construct a `closed time path' action
principle\cite{schwinger2,bakshi1}.

Consider an expectation value of the form
\begin{equation}
\langle t| X(t')|t\rangle = \sum_{i,i'} \langle t|i\rangle\langle i|X(t')|i'
\rangle\langle i'|t \rangle\label{eq:14}
\end{equation}
where the sum over intermediate states $i,i'$ is a sum over all states
and $\langle t|$ is a shorthand which refers to either a pure state of the
system, or a
mixed state, specified at time $t$. The expectation value specifies
the average value of the operator $X$ at the time $t'$ given the state
of the system at time $t$. It involves conjugate amplitudes and hence the
conjugate forms of the
action principle:
\begin{eqnarray}
\delta \langle t|i\rangle &=& i\langle t|\delta S_{ti}|i \rangle\label{eq:15}\\
\delta \langle i|t \rangle &=& -i\langle i|\delta S_{it}^\dagger|t
\rangle\label{eq:16}\\
S_{ab} &=& \int_a^b dt L\label{eq:17}.
\end{eqnarray}
To obtain (\ref{eq:14}) from an action principle one would therefore like to
introduce the operator $X$ by functional differentiation with respect to an
appropriate source (or combination of sources) between an amplitude and its
conjugate. This is achieved in the following way. First one observes that
\begin{equation}
\delta \langle t|t \rangle = \delta(\sum_i \langle t|i \rangle \times \langle
i|t \rangle)
= \delta(1) = 0\label{eq:18},
\end{equation}
so differentiation of this object is to no avail.
However, if we make an artificial distinction between the amplitude and its
conjugate by labelling all objects in the former with a $+$ symbol and all
objects in the latter with a $-$ symbol,
\begin{equation}
\delta \langle t|t\rangle = \lim_{+\rightarrow -}i \langle t|\delta S_+-\delta
S_-^\dagger|t\rangle
\label{eq:19}
\end{equation}
then we can use the solution of this quantity as a generating functional
for (\ref{eq:14}) since $X$ can be expanded in terms of either
$\frac{\delta}{\delta J_+}$
or$\frac{\delta}{\delta J_-}$. This breaks the symmetry of symbols in
(\ref{eq:18}).
At the end of a variational calculation one removes the $+$ and $-$ symbols
restoring the conjugate relationship between the two amplitudes,
having inserted the appropriate operators by differentiation with respect
to the source of only one of them. Note in (\ref{eq:19})
that, for any unitary field theory, the action is self-adjoint, thus we may
drop the dagger symbol in future. Also, in treating the $+$ and $-$ parts of
the
field as being artificially independent, the condition
\begin{equation}
\phi_+(t_{\infty}) = \phi_-(t_{\infty})\label{eq:20}
\end{equation}
is required to ensure that the limit $+\rightarrow -$ restores the single
identity of the field operators, and additionally one must have that all
$\phi_-$ fields (at any time) must stand to the left of all $\phi_+$ fields
(at any time). Since $-$ fields are anti-time-ordered and $+$ fields
are time ordered, this condition arises naturally and ensures the triviality
of (\ref{eq:13}).

The meaning of the above procedure can be illustrated by noting that the
solution
to (\ref{eq:19}) may be written
\begin{equation}
\langle t_i|t_i\rangle_{J_{\pm}} = \langle t_i|T^\dagger e^{-iJ_i\phi_-}T
e^{iJ_+\phi_+}  |t_i\rangle
\label{eq:21}.
\end{equation}
The expectation value of the field is found using the ordered expression
\begin{eqnarray}
-i\frac{\delta}{\delta J_+(x)} \langle t_i|t_i\rangle \Bigg |_{+=-} &=& \langle
\phi(x) \rangle\\
&=& \lim_{+\rightarrow -}\langle t_i|
\exp\left(-i\int_{t_i}^{\infty}J_-\phi_-
+i\int_{t}^{\infty}J_+\phi_+ \right)
\phi_+(x)
\exp\left(i\int_{t_i}^{t}\right)
|t_i\rangle\label{eq:22}.
\end{eqnarray}
Taking the limit $+\rightarrow -$, one has
\begin{equation}
\langle \phi(x)\rangle = \langle t_i|\exp\left(-i\int_{t_i}^{t}J\phi\right)
\phi(x)
\exp\left(i\int_{t_i}^{t}J\phi\right)
|t_i\rangle.
\label{eq:23}
\end{equation}
This shows that the average value of the operator depends only on the
past (retarded) history of the system beginning from the initial time $t_i$.
It can be shown (see appendix A) that the closed time path generating
functional is closely related to the generator for the retarded $n$-point
functions.
The acausal (advanced) pieces cancel in the limit $+\rightarrow -$.

So far, the discussion has used the slightly trivial example of pure states
$\langle t|$.
As noted implicitly by Schwinger\cite{schwinger2}, the same action principle
holds when $\langle t| \ldots |t\rangle$ is replaced by $\langle t|\rho(t)
\ldots |t\rangle$
(a mixture of states) since this does not affect
the conjugate relationship between amplitudes.
The nature of the expectation value can therefore be left out
of the discussion for the most part. Indeed, in practice, the effect of a
non-trivial
density matrix in the expectation value can be mimicked by the introduction of
suitable sources\cite{schwinger2,calzetta2}---a procedure which will be adopted
in the next section. To present the formalism in a way conducive to
generalization, the next step is to present the Green functions for the
case of pure-state vacuum expectation values and then introduce the finite
temperature (mixed state) modifications which will be the starting point for
writing down
an ansatz for non-equilibrium.

The above use of generating functionals is closely related to the path integral
approaches of Calzetta and Hu\cite{calzetta1}, and Lawrie\cite{lawrie1}.
It proves useful not to pass directly to the path integral however, but to
follow Schwinger's approach.
For the remainder of the paper, equation (\ref{eq:19}) will be considered the
starting point for the discussion of non-equilibrium field theory.

{}From equations (\ref{eq:1}) and (\ref{eq:19}) one obtains the operator
equations
of motion for the field. Taking the initial time to be
$t_i$, the furthermost future time to be $t_\infty$ and
the final time at which expectation values are to be computed
as $t_f$, then using the boundary condition in equation (\ref{eq:20}),
\begin{eqnarray}
\phi_+(x) &=& \phi(x_i) + \int_{t_i}^{t} G_c(x,x')J_+(x') dV_{x'}\nonumber\\
\phi_-(x) &=& \phi_+(t_\infty) + \int_{t_\infty}^{t_f} G_c(x,x') J_-(x')
dV_{x`}\nonumber\\
          &=& \phi(x_i) + \int_{t_i}^{t_\infty} G_c(x,x')J_+(x') dV_{x'}+
\int_{t_\infty}^{t_f} G_c(x,x') J_-(x') dV_{x`}\label{eq:24}
\end{eqnarray}
where $G_c(x,x')$ is a retarded (causal) Green function.
Notice that, as the distinction between $+$ and $-$ is removed, these equations
reduce to (\ref{eq:4}). Substituting these into the exponential solution to
(\ref{eq:19}) and defining a vector and its transverse by $J^T = (J_+ J_-)$,
one
may write
\begin{equation}
\ln \langle 0,t_i|0,t_i\rangle = -i\int_{t_f}^{t_\infty}
J_-(x')dV_{x'}\phi(x_i)
+ i \int_{t_i}^{t_\infty} J^T(x) G(x,x') J(x') dV_{x}dV_{x'}
\label{eq:25}
\end{equation}
where
\begin{equation}
G(x,x') = \left(
\begin{array}{cc}
\theta(x-x')G_c(x,x') & 0 \\
-G_c(x,x') & \theta(x'-x)G_c(x,x')
\end{array}
\right),
\label{eq:26}
\end{equation}
and $\theta(x-x')$ is the step function which satisfies
\begin{equation}
\theta(x) + \theta(-x) = 1\label{eq:27}.
\end{equation}
As a result of this property, the sum of rows and columns in (\ref{eq:26}) is
zero. This is a reflection of the triviality of equation (\ref{eq:13}).
It further implies the causality of expectation values derived from this
generating functional. While (\ref{eq:26}) has a simple physical
derivation in terms of the equations of motion, a more symmetrical
form can be obtained by attaching a variational interpretation to
$G_c(x,x')$ directly. Again, following Schwinger, and varying with respect
to the sources
\begin{equation}
\delta_2\delta_1 \langle t_i|t_i \rangle = (i)^2 \langle
t|(\delta_2S_+-\delta_2S_-)(\delta_1S_+-\delta_1S_-) |\rangle t>
\label{eq:28}
\end{equation}
where, according to the ordering rule, this equals
\begin{equation}
\delta_2\delta_1 \langle t_i,t_i \rangle = (i)^2 \langle
t|\phi_+(x_2)\phi_+(x_1)
+\phi_-(x_2)\phi_-(x_1) - \phi_-(x_2)\phi_+(x_1) - \phi_+(x_2)\phi_-(x_1)
|\rangle t>.
\label{eq:29}
\end{equation}
Comparing the solution of this to
\begin{equation}
\exp\left( \frac{i}{2}\int dV_x dV_{x'} J^T G(x,x') J(x')\right) \label{eq:30}
\end{equation}
one has
\begin{equation}
G(x,x') = \left(
\begin{array}{cc}
G_{++} & G_{+-} \\
G_{-+} & G_{--}
\end{array}
\right),
\label{eq:31}
\end{equation}
where
\begin{eqnarray}
\langle \phi_+(x)\phi_+(x')\rangle &=& -iG_{++}(x,x')\label{eq:32}\\
\langle\phi_+(x)\phi_-(x') \rangle &=& iG_{+-}(x,x')\label{eq:33}\\
\langle\phi_-(x)\phi_+(x') \rangle &=& iG_{-+}(x,x')\label{eq:34}\\
\langle \phi_-(x)\phi_-(x')\rangle &=& -iG_{--}(x,x')\label{eq:35}.
\end{eqnarray}
As the distinction between $+$ and $-$ is lifted, the assumed ordering
implies that
\begin{eqnarray}
G_{++}(x,x') &=& i \langle T \phi(x)\phi(x')\rangle  = G_F(x,x')\label{eq:36}\\
G_{+-}(x,x') &=& -i \langle \phi(x)\phi(x')\rangle =
-G^{(-)}(x,x')\label{eq:37}\\
G_{-+}(x,x') &=& -i \langle\phi(x')\phi(x)\rangle =
G^{(+)}(x,x')\label{eq:38}\\
G_{--}(x,x') &=& i \langle T^\dagger\phi(x)\phi(x') \rangle =
G_{AF}(x,x')\label{eq:39}\\
\end{eqnarray}
where $G_F$ is the Feynman propagator, $G^{(\pm)}$ are the positive and
negative
frequency Wightman functions and $G_{AF}$ is the anti-time ordered propagator.
In the limit of zero source, these quantities satisfy the equations
\begin{eqnarray}
(-{\vcenter{\vbox{\hrule height.4pt\hbox{\vrule width.4pt
height8pt\kern8pt\vrule width.4pt}\hrule height.4pt}}} + m^2) G_F(x,x') &=&
\delta(x,x')\label{eq:40}\\
(-{\vcenter{\vbox{\hrule height.4pt\hbox{\vrule width.4pt
height8pt\kern8pt\vrule width.4pt}\hrule height.4pt}}} + m^2) G^{(\pm)}(x,x')
&=& 0\label{eq:41}\\
(-{\vcenter{\vbox{\hrule height.4pt\hbox{\vrule width.4pt
height8pt\kern8pt\vrule width.4pt}\hrule height.4pt}}} +m^2) G_{AF}(x,x') &=& -
\delta(x,x')\label{eq:42}\\
\end{eqnarray}
by virtue of equation (\ref{eq:5}) for the field operator. The
non-zero right hand side of (\ref{eq:40}) and (\ref{eq:41}) are
due to the time ordering. From the time-ordering (\ref{eq:36}) it
follows that
\begin{equation}
-iG_F(x,x') = i\theta(t-t')G^{(+)}(x,x') -i\theta(t'-t)G^{(-)}(x,x').
\label{eq:43}
\end{equation}
Substituting this relation into (\ref{eq:36}) and using the commutation
relations for the field (\ref{eq:9}) proves (\ref{eq:36}).
Similarly (\ref{eq:42}) follows from the relation
\begin{equation}
G_{AF}(x,x') = - G_F^*(x,x')\label{eq:44}.
\end{equation}
Using (\ref{eq:27}), it is now straightforward to see that the sum
of the rows and columns in (\ref{eq:31}) is vanishing, as required
for causality. A number of additional relations between the
Green functions can be proven. The retarded and advanced Green
functions satisfy:
\begin{eqnarray}
G_{ret}(x,x') &=& -\theta(t-t')\lbrack \phi(x),\phi(x')\rbrack\nonumber\\
G_{adv}(x,x') &=& \theta(t'-t)\lbrack \phi(x),\phi(x')\rbrack\label{eq:45}.
\end{eqnarray}
Also, in virtue of (\ref{eq:27}) it is easy to see that
\begin{eqnarray}
G_F &=& G_{ret} + G^{(-)} = G_{adv} - G^{(+)}\label{eq:46}\\
G_{AF} &=& = G_{ret}+G^{(+)} = G_{adv} - G^((-))\label{eq:47}.
\end{eqnarray}
The unequal-time commutator and anti-commutator Green-functions are defined
by
\begin{eqnarray}
\tilde{G}(x,x') &=& \lbrack \phi(x),\phi(x')\rbrack =
G^{(+)}+G^{(-)}\nonumber\\
\overline{G}(x,x') &=& \lbrace \phi(x),\phi(x')\rbrace = G^{(+)}-G^{(-)}
\label{eq:48}
\end{eqnarray}
These will be useful later and serve to pin-point the conventions used
in this work. Before considering the momentum-space
representation of these functions it is useful to note
that $G(x,x')$ can be written entirely in terms of the formal quantity
\begin{equation}
H(x,x') \equiv i \langle t|\phi(x)\phi(x')|t\rangle\label{eq:49}
\end{equation}
as
\begin{equation}
G(x,x') = \theta(t-t')\left(
\begin{array}{cc}
H(x,x') & -H(x',x) \\
-H(x,x') & H(x',x)
\end{array}
\right)
+
\theta(t'-t)
\left(
\begin{array}{cc}
H(x',x) & -H(x',x) \\
-H(x,x') & H(x,x')
\end{array}
\right)
,
\label{eq:50}
\end{equation}
where
\begin{equation}
H(x',x) = H(x,x')^*
\label{eq:51}
\end{equation}
and the sum of rows and columns is manifestly zero.
Since the spectrum of the operator $-{\vcenter{\vbox{\hrule
height.4pt\hbox{\vrule width.4pt height8pt\kern8pt\vrule width.4pt}\hrule
height.4pt}}}+m^2$ on the
complex wave $e^{ikx}$ is solved for any $k$ satisfying
a dispersion relation, the solution to (\ref{eq:41}) is
the most general linear combination of plane waves satisfying
the dispersion relation $k^2+m^2=0$. This implies that the vacuum
positive and negative frequency Wightman functions can be
written, in $n$ spacetime dimensions,
\begin{eqnarray}
G^{(+)} &=& - 2\pi i \int \frac{d^nk}{(2\pi)^n}e^{ik_\mu(x-x')^\mu}
\theta(k_0)\delta(k^2+m^2)\label{eq:52}\\
G^{(-)} &=& 2\pi i \int \frac{d^nk}{(2\pi)^n}e^{ik_\mu(x-x')^\mu}
\theta(-k_0)\delta(k^2+m^2)\label{eq:53}.
\end{eqnarray}
Defining the fourier transform of $G(x,x')$ by
\begin{equation}
\int \frac{d^nk}{(2\pi)^n} e^{ik_\mu(x-x')^\mu} G(k),\label{eq:54}
\end{equation}
and using the integral representation
\begin{equation}
\theta(t-t') = i \int_{-\infty}^{+\infty} \frac{d\omega}{2\pi}
 \frac{e^{-i\omega(t-t')}}{\omega+i\epsilon}\label{eq:55}
\end{equation}
it is straightforward to show from (\ref{eq:43}) that
\begin{equation}
G_F(k) = \frac{1}{k^2+m^2-i\epsilon},\label{eq:56}
\end{equation}
which is fully consistent with (\ref{eq:40}). $G_{AF}(k)$ is
easily obtained from (\ref{eq:44}).

Note that, had the symmetrical form of $G(x,x')$ not been used,
similar results could still have been obtained. It is possible, in
the manner of a symmetry transformation to redefine the
Wightman functions so that positive and negative frequencies are
mixed. This simply mixes up the Feynman and anti-Feynman propagator
also. For instance, if one defines
\begin{equation}
G^{(+)}(k) = 2\pi i\delta(k^2+m^2) \lbrack\theta(k_0) +\alpha\theta(k_0)
+\beta\theta(-k_0)\rbrack
\label{eq:57},
\end{equation}
then the corresponding Feynman propagator becomes
\begin{equation}
G_F(k) = \frac{1+\alpha}{k^2+m^2-i\epsilon} -
\frac{\beta}{k^2+m^2+i\epsilon}\label{eq:58}
\end{equation}
where the last term is evidently a piece from $G_{AF}$.
Since this only complicates the matter, such redefinitions will not
be pursued further.

So far, this summary of the action principle has only
explicitly encompassed pure state
expectation values, which are comparatively trivial. A statistical system
with real particle densities, as well as perhaps a temperature and entropy,
is described by a mixture of such vacuum expectation values (since the
character of the actual pure state is not usually knowable), with the
the statistical weight given by the density matrix $\rho$. The simplest
example of such is a system in thermal equilibrium ($\rho = \exp(-\beta H)$).
Although a thermal system is quite extraordinary as many particle systems
go, it serves as a useful reference point, both from the viewpoint of
formalism and from a physical perspective, since very many physical systems
can be characterized by a temperature of sorts. A statistical
expectation value for some operator $X$ may be written
\begin{equation}
\langle t| X(t')|t \rangle_s \equiv \frac{{\rm Tr} \langle t| \rho(t) X(t')  |t
\rangle}
{{\rm Tr} \langle t| \rho(t) |t \rangle}
\label{eq:59}
\end{equation}
and characterizes the average value of $X$ at the time $t'$ given the
state of the system at time $t$. Notice that the trace is over
{\em probabilities} of the form $\langle t|t \rangle$ rather than
amplitudes $\langle t|t'\rangle$. The latter would be meaningless.
The structure of the expectation value is therefore simply that in
equation (\ref{eq:14}) and the closed time path action principle
applies. Indeed, it is noteworthy that the density matrix itself
is merely an operator which can effectively by introduced into
the pure state generating functional by functional differentiation
with respect to an appropriate source. There is therefore no loss
of generality in taking the closed time path action principle
at face value and making no special reference to $\rho$..

The cyclic property of the trace in (\ref{eq:59}) has noteworthy
implications for the Green functions and sources in the CTP formalism.
Consider the expectation value in (\ref{eq:59}). This may be
rewritten as
\begin{eqnarray}
\langle X(t')\rangle_s &=&
\frac{{\rm Tr}\langle \rho(t) e^{iH(t'-t)}X(t)e^{-iH(t'-t)}\rangle}
{{\rm Tr} \langle t| \rho(t) |t \rangle}\nonumber\\
 &=&  \frac{{\rm Tr}\langle e^{-iH(t'-t)} \rho(t) e^{iH(t'-t)}X(t) \rangle}
{{\rm Tr} \langle t| \rho(t) |t \rangle}
\label{eq:60}
\end{eqnarray}
where $H$ is the Hamiltonian of the system.
Using this `relativity' between the time-dependence of $\rho$ and $X$, it
is possible to place all of the dynamical development of the system
in either one or the other. An example of the use of density matrix
time-development is given in ref. \cite{eboli1}. In the CTP formalism,
the distinction between forward moving times and backward moving
times makes equation (\ref{eq:60}) effectively
\begin{equation}
\langle X(t')\rangle_s=
\frac{\langle e^{-iH_-(t'-t)} \rho(t) e^{iH_+(t'-t)}X(t) \rangle}
{ {\rm Tr} \langle t | \rho(t) |t \rangle}.
\label{eq:61}
\end{equation}
The cyclic property of the trace therefore implies that the density
matrix $\rho$ always sits between the $+$ and $-$ branches of the
operator product and hence it must always be reflected by the
{\em off-diagonal} terms in $\pm$-space. In the special case of
a thermal density matrix, the same observation leads to the well-known
KMS condition\cite{martin1}, by identifying the inverse-temperature $\beta$
with imaginary time. This is seen by considering the thermal Wightman function
\begin{eqnarray}
G^{(+)}_\beta(x,x') &=& \frac{{\rm Tr} \langle t| e^{-\beta H}
\phi(x)\phi(x')|t \rangle}
{{\rm Tr} \langle t| e^{-\beta H} |t \rangle}\nonumber\\
&=& \frac{{\rm Tr} \langle t| e^{-\beta H} \phi(x)e^{\beta H}e^{-\beta
H}\phi(x')|t \rangle}
{{\rm Tr} \langle t| e^{-\beta H} |t \rangle}\nonumber\\
&=&\frac{{\rm Tr} \langle t| e^{-\beta H} \phi(x')\phi({\bf x},t+i\beta)|t
\rangle}
{{\rm Tr} \langle t| e^{-\beta H} |t \rangle}\nonumber\\
&=& - G^{(-)}({\bf x},t+i\beta,x')
\label{eq:62}
\end{eqnarray}
where the cyclic property of the trace has been used.
The left and right hand sides are precisely the elements
of the off-diagonal $G_{+-}$ and $G_{-+}$.
This property is, in fact, sufficient to determine the
thermal Green functions.

To determine these, in a form which manifestly reduced to
the vacuum case, one writes
\begin{eqnarray}
G^{(+)}(k) = -2\pi i \lbrack\theta(k_0)+X \rbrack\delta(k^2+m^2)\nonumber\\
G^{(-)}(k) = 2\pi i \lbrack\theta(-k_0)+Y \rbrack\delta(k^2+m^2)
\label{eq:63}
\end{eqnarray}
with $X$ and $Y$ to be determined. Since the commutator
function $\tilde{G}(x,x')$ must be independent of the state
of the system (in order to preserve the canonical commutation
relations), it follows immediately that $X=Y$. If one then employs
the KMS condition which, in momentum space, becomes the definite relation
\begin{equation}
G^{(+)}(k) = - e^{\beta k_0}G^{(-)}(k)
\label{eq:64}
\end{equation}
it follows that
\begin{equation}
\theta(k_0) +X = e^{|k_0|\beta} \lbrack \theta(-k_0)+X\rbrack
\label{eq:65}
\end{equation}
and hence
\begin{equation}
X = \theta(k_0)f(|k_0|) = f_{>0}(k_0)
\label{eq:66}
\end{equation}
where
\begin{equation}
f(|k_0|) = \frac{1}{e^{\beta|k_0|}-1}.
\label{eq:67}
\end{equation}
By considering (amongst other things) $\overline{G}(x,x')$, it
follows that $f(|k_0|)$ is an even function of $|k_0|$ thus
$f(k_0)\theta(k_0)=f(-k_0)\theta(-k_0)$, whereupon it is
trivial to show the unitarity relation
\begin{equation}
G^{(+)}(x,x') = G^{(-)}(x,x')^*.
\label{eq:68}
\end{equation}
Note that the fact that $G^{(+)}$ consists only of positive
frequencies $k_0$ is pivotal in this derivation.  The Feynman
propagator can now be obtained from equation (\ref{eq:43}) by using
the integral representation of the step-function (\ref{eq:55}). The
thermal Green functions are therefore summarized by
\begin{eqnarray}
G^{(+)}(k) &=& -2\pi i \theta(k_0) \lbrack 1+f(|k_0|) \rbrack
\delta(k^2+m^2)\label{eq:69}\\
G_F(k) &=& \frac{1}{k^2+m^2-i\epsilon} +2\pi i
f(|k_0|)\delta(k^2+m^2)\theta(k_0)\label{eq:70}.
\end{eqnarray}

Another important form of $G^{(\pm)}$ is obtained by
performing the integral over $k_0$, thereby enforcing the
role of the delta-function in (\ref{eq:63}). This gives
a result which will prove more useful for calculations later and is more
closely
related to the ansatz used by Lawrie in ref. \cite{lawrie1}:
\begin{eqnarray}
G^{(+)}(x,x')&=& -2\pi i \int \frac{d^{n-1}k}{(2\pi)^{n-1}}
e^{i({\bf k\cdot} ({\bf x}-{\bf x'})-\omega (t-t'))}
\frac{(1+2f_{>0}(|\omega|))}{2|\omega|}\nonumber\\
&=& -2\pi i \int \frac{d^{n-1}k}{(2\pi)^{n-1}}
e^{i({\bf k\cdot} ({\bf x}-{\bf x'})-\omega (t-t'))}
\frac{(1+f(|\omega|))}{2|\omega|}
\label{eq:71}
\end{eqnarray}
where $f_{>0}(k_0)$ is the function composed of only positive
frequencies.
It is now straightforward to verify that the canonical commutation
relations are satisfied, by differentiating $\tilde{G}(x,x')$
with respect to $t'$ (see equation (\ref{eq:48})).

This completes the presentation of conventions
to be used in the remainder of the paper. It is convenient to
add here that a bar (e.g. $\overline a$) represents
an object which is even, while an object with a tilde (e.g. $\tilde a$)
represents one which is odd with respect to its arguments.

\section{Interaction with sources}

The formalism demonstrated so far has been for free fields. Free
fields are always in a state of equilibrium and therefore the discussion needs
to be
widened to incorporate collisions or interactions. The present work
will deal with interactions which can be mediated by sources of the
type $\phi(x)A(x,x')\phi(x')$. This includes a variety of self interactions,
contact with external forces and noise or impurity scattering,
depending on the nature of $A(x,x')$.
The self-energy of an interacting field theory has this form,
for instance, thus sources of the quadratic type can also be a representation
of the
lowest order, self-consistent `particle dressings'. In Feynman diagram
language,
these represent the re-summed one-loop approximation, Hartree approximation
and so on. Lawrie\cite{lawrie1} uses the notion of such sources to effect
a renormalization (resummation) of perturbation theory in a real scalar
field theory. The same idea is expressed in a different language in the
work of Calzetta and Hu\cite{calzetta1}.
Since it is not the aim of this paper to discuss specific
models, the specific nature of the source terms will not be specified here.
Rather
the discussion will centre around what general properties such a system
might have and a discussion of possible applications will follow.

In an interacting theory, one normally perturbs about the free
field theory. Unfortunately, the dispersion relation (or `mass shell'
constraint) for
free particles is no longer appropriate, since it reflects none of the
interactions which `dress' the particles. A more satisfactory starting
point would be a `quasi-particle' mass shell, including some of the
interaction effects as the basis for a perturbation theory.
This is the essence of a renormalization and can be effected by the
use of sources\cite{burgess99}.

The starting point for the investigation of non-equilibrium fields will
therefore be the closed time path (CTP) action principle, taking the action $S$
for a neutral scalar boson and supplementing it by quadratic sources.
Observing the CTP operator ordering, one has
\begin{eqnarray}
S_+-S_- \rightarrow S_+-S_- &+&\frac{1}{2}\int dV_x dV_{x'} \lbrack
T(\phi_+(x)A_{++}(x,x')\phi_+(x')) \nonumber\\
&+& \phi_-(x)A_{-+}(x,x')\phi_+(x')\nonumber\\
&+& \phi_-(x')A_{+-}(x,x')\phi_+(x)\nonumber\\
 &+& T^\dagger(\phi_-(x)A_{--}(x,x')\phi_-(x'))
\rbrack
\label{eq:72}
\end{eqnarray}

It should be clear that no fundamental field theory may contain
off-diagonal terms in $\pm$-space. The CTP action principle is, by
construction,
diagonal, being the difference between $S_+$ and $S_-$ (see equation
(\ref{eq:19})).
However, it was remarked earlier that the effect of a density matrix must
be reflected in off-diagonal terms so, while such terms are certainly
not fundamental, they can exist as off-diagonal self-energies representing
the dynamics of a density matrix. Moreover, since off-diagonal terms represent
a point of contact between fields moving forward in time and fields moving
backward in time, one might anticipate that off-diagonal sources would be
at least partly responsible for choosing an arrow of time (the generation
of entropy). The explicit coupling will therefore play an important role
in both non-equilibrium kinematics and dynamics.

The essential unitarity of the CTP formalism is seen, from equations
(\ref{eq:15})
and (\ref{eq:16}), to be summarized by the
following property of the transformation function:
\begin{equation}
\langle t_i | t_i\rangle^*_{\pm} = \langle t_i | t_i\rangle_{\mp}\label{eq:73}
\end{equation}
namely that complex conjugation merely exchanges $+$ labels with $-$ labels
and vice versa.
If one defines indices $a,b = +,-$, then the operator defined by the
second variation of (\ref{eq:72}) with respect to the field $\phi_a$,
$S_{ab}=\delta_a\delta_b(S_+-S_-)$, with
$S_{ab} = S_{++}, S_{+-}\ldots$, satisfies the relations
\begin{eqnarray}
S_{++}^*(x,x') &=& -S_{--}(x,x')\nonumber\\
S_{+-}^*(x,x') &=& - S{-+}(x,x').
\label{eq:74}
\end{eqnarray}
This, in turn, implies that $S_{ab}$ may be written
in terms of real constants $A,B,C$ and $\gamma_\mu$.
\begin{eqnarray}
S_{ab}(x,x') =
\left(
\begin{array}{cc}
(-{\vcenter{\vbox{\hrule height.4pt\hbox{\vrule width.4pt
height8pt\kern8pt\vrule width.4pt}\hrule
height.4pt}}}+m^2)\delta(x,x')+A(x,x')+iC(x,x') & B(x,x') + \gamma^\mu(x,x')
\stackrel{\leftrightarrow}{D^\gamma_{\mu}}' -iC(x,x')\\
-B(x,x') - \gamma^\mu(x,x') \stackrel{\leftrightarrow}{D^\gamma_{\mu}}'
-iC(x,x') & ({\vcenter{\vbox{\hrule height.4pt\hbox{\vrule width.4pt
height8pt\kern8pt\vrule width.4pt}\hrule
height.4pt}}}-m^2)\delta(x,x')-A(x,x')+iC(x,x')
\end{array}
\right)
\label{eq:75}
\end{eqnarray}
where a new derivative has been defined to commute with the function
$\gamma^\mu(x,x')$:
\begin{equation}
\stackrel{x}{D_\mu^\gamma}~ \equiv ~\stackrel{x}{\partial}_\mu +
\frac{1}{2}\gamma_\nu^{-1}(x,x')\stackrel{x}{\partial}_\mu\gamma_\nu(x,x').
\label{eq:76}
\end{equation}
Note, first of all, that the sum of rows and columns in this operator
is zero, as required for unitarity and subsequent causality.
Derivatives higher than first order in the sources could be rewritten
using the field equations (still to be found) and absorbed into other
terms, thus such terms are redundant. There can be no non-vanishing
terms of the form $\phi_+\partial\phi_+$ without violating time
reversal invariance or merely adding total derivatives to the
action. Finally $C\not=0$ is clearly disallowed in a fundamental
theory on the grounds of unitarity. It turns out, by considering the
field equations, that the only fully consistent choice is $C=0$, even
though such a term does not violate equation (\ref{eq:73}).  Equation
(\ref{eq:75}) agrees with the form given by Lawrie, up to differences
in conventions and the inclusion here of $B(x,x')$.

The significance of the off-diagonal terms involving $\gamma^\mu$ can be seen
by writing out the coupling fully:
\begin{equation}
\gamma^\mu(x,x') \cdot \left( \phi_1 D^\gamma_\mu \phi_2 - \phi_2 D^\gamma_\mu
\phi_1\right).
\label{eq:gamma}
\end{equation}
The term in parentheses has the form of a current between components $\phi_1$
(the the forward moving field) and $\phi_2$ (the backward moving field).
When these two are in equilibrium there will be no dissipation to the
external reservoir and these
off-diagonal terms will vanish. This indicates that these off-diagonal
components
(which are related to off-diagonal density matrix elements, as noted earlier)
can be
understood as the mediators of a detailed balance condition for the field.
A similar conclusion was reached in reference \cite{calzetta1} by a different
argument for the quantity referred to here as $\tilde B$. When the term is
non-vanishing, it represents a current flowing in one particular direction,
pointing out the arrow of time for either positive or negative frequencies.
The current is a `canonical current' and is clearly related to the fundamental
commutator for the scalar field in the limit $+\rightarrow -$.

Using equation (\ref{eq:75}) it is now possible to express (\ref{eq:72})
in the form
\begin{equation}
S_{CTP} = \int dV_x dV_{x'} \frac{1}{2}\phi^a S_{ab} \phi^b
\label{eq:77}
\end{equation}
and thus the closed time path field equations may be found by varying
this action with respect to the $+$ and $-$ fields:
\begin{eqnarray}
\frac{\delta S_{CTP}}{\delta\phi_+(x)} &=&  (-{\vcenter{\vbox{\hrule
height.4pt\hbox{\vrule width.4pt height8pt\kern8pt\vrule width.4pt}\hrule
height.4pt}}}+m^2)\phi_+(x)
+\frac{1}{2}\int dV_{x'}
\Bigg\lbrace
(\overline A+i\overline C)\phi_+(x') + (\tilde{B} -i\overline C)\phi_-(x')
+\overline \gamma^\mu(\partial_\mu\phi_-(x'))
+\overline{\partial^\mu\gamma_\mu}\phi_-(x')
\Bigg\rbrace\label{eq:78}\\
\frac{\delta S_{CTP}}{\delta\phi_-(x)} &=&  ({\vcenter{\vbox{\hrule
height.4pt\hbox{\vrule width.4pt height8pt\kern8pt\vrule width.4pt}\hrule
height.4pt}}}-m^2)\phi_-(x)
-\frac{1}{2}\int dV_{x'}
\Bigg\lbrace
(\overline A-i\overline C)\phi_+(x') + (\tilde{B} +i\overline C)\phi_-(x')
+\overline \gamma^\mu(\partial_\mu\phi_+(x'))
+\overline{\partial^\mu\gamma_\mu}\phi_+(x')
\Bigg\rbrace\label{eq:79}\\
\end{eqnarray}
and setting the right hand side to zero,
which introduces the notation
\begin{eqnarray}
\overline A(x,x') &=& \frac{1}{2}\left(A(x,x')+A(x',x) \right)\nonumber\\
\tilde{A}(x,x') &=& \frac{1}{2}\left(A(x,x')-A(x',x) \right)\nonumber\\
\overline{\partial^\mu\gamma_\mu}(x,x') &=&  \frac{1}{2}\left(
\stackrel{x}{\partial^\mu}
\gamma_\mu(x,x')+ \stackrel{x'}{\partial^\mu}
\gamma_\mu(x',x)
\right)
\label{eq:80}.
\end{eqnarray}
The canonical commutation relations which derive from the action
principle (see equation (\ref{eq:9})) are unchanged by these
modifications, since they depend only on $\overline x$
and therefore cancel in the commutator.  This is key feature in any
consistent description of non-equilibrium phenomena.

To solve this system of non-local equations, the best strategy will
be to look for the Green functions, or the inverse of the operator
$S_{ab}$. This is the method adopted by Lawrie\cite{lawrie1}.
Although a common strategy will be used here, the method will
be somewhat different in spirit. The variational approach
used in ref. \cite{calzetta1} will not be used here.
Owing to the non-locality, it is clear that the inverse
operator cannot be a translationally invariant function in the
general case. It must be formally dependent on both Cartesian
coordinate differences and the average coordinate:
\begin{eqnarray}
\tilde{x} \equiv \frac{1}{2}(x-x')\label{eq:81}\\
\overline{x} \equiv \frac{1}{2}(x+x')\label{eq:82}.
\end{eqnarray}
Moreover, since the operator contains off-diagonal terms,
which typically signify a non-trivial density matrix, it
is natural to look for a solution based on the form of
equation (\ref{eq:69}), generalized to include a dependence
on $\overline x$. Although this sounds like an
innocent step, it is far from a trivial undertaking since it
introduces non-linearities in the spectrum of excitations which
must be handled in a self-consistent way. It is useful to
work with the quantity $H(x,x')$, from which all the Green
functions can be obtained (actually the Wightman
function in disguise). Using either the field equations or
the matrix equation
\begin{equation}
S_{ab}G^{bc} = \delta_a^{~c}\delta(x,x')\label{eq:83}
\end{equation}
one obtains equations of motion for the quantity $H(x,x')$ (see
equation (\ref{eq:50})).  Not all of these equations are independent,
owing to the symmetry in equation (\ref{eq:73}). In particular, their
consistency requires that $C=0$ which is now chosen explicitly.  It is
sufficient to consider
\begin{equation}
(-{\vcenter{\vbox{\hrule height.4pt\hbox{\vrule width.4pt
height8pt\kern8pt\vrule width.4pt}\hrule height.4pt}}}+m^2)H(x,x') + \int
dV_{x''} \left(\overline A(x,x'')
-\tilde{B}(x,x')-\overline\gamma^\mu(x,x'')\stackrel{x''}{\partial}_\mu
-\overline{\partial_\mu\gamma^\mu}(x,x'')\right)H(x'',x) = 0
\label{eq:84}
\end{equation}
on the understanding that $H(x,x') = H(x',x)^*$.
This relation is to be supplemented by the canonical
commutation relations for the field, which appear in
equation (\ref{eq:83}) in the form
\begin{equation}
\partial_{t'} \left( H(x,x')-H(x',x) \right)\Bigg|_{t=t'}=i\delta({\bf x},{\bf
x'})
\label{eq:85}
\end{equation}
and complete the consistency of the equations of motion.

The next step in obtaining an intuitive formalism is to introduce
a (local) momentum space technique by Fourier transforming
$\tilde x$ and retaining a dependence on the average coordinate
$\overline x$:
\begin{equation}
H(x,x') = \int \frac{d^nk}{(2\pi)^n} e^{ik(x-x')}H(k,\overline x)\label{eq:86}.
\end{equation}
A suitable ansatz for this function, which generalizes the
dispersion relation and the one-particle distribution function $f(k_0)$
is
\begin{equation}
H(k,\overline x) = 2\pi \theta(k_0)\lbrack 1+f(k_0,\overline x)\rbrack
\delta(-k_0^2+\omega^2(k,\overline x))
\label{eq:87}.
\end{equation}
The spacetime dependent function $f(k,\overline x)$ is often referred to
as the Wigner function and signifies the inhomogeneity in particle
occupation numbers.
The generalized dispersion relation takes generic form
$-k_0^2+\omega^2 =0$. In the free particle limit $\omega^2={\bf k}^2+m^2$.
It is the determination of this dispersion relation which is
of specific importance, since this determines the spectrum of
excitations for particles in the plasma-field, and forms the
basis of all perturbation theory when the sources represent
self-interactions.

It can be verified that, since $H(k,\overline x)$ depends only on
the average coordinate, the commutation relations are preserved (see equation
(\ref{eq:85})).
A more useful form of (\ref{eq:86}) is obtained on performing the integration
over $k_0$. This eliminates the dubious derivative of the delta-function
from subsequent relations and leads to a number of helpful insights.
\begin{equation}
H(x,x') = 2\pi \int \frac{d^{n-1}k}{(2\pi)^{n-1}}e^{ik_\mu(x-x')^\mu}
\frac{(1+f(k_0,\overline x))}{2|\omega|}
\label{eq:88}
\end{equation}
where it is understood that $k_0=|\omega|$.
Finally, it is useful to define the derivative with respect
to the average coordinate $\overline \partial =
\frac{1}{2}(\partial_x+\partial_{x'})$
and the quantities
\begin{eqnarray}
F_\mu &=& \frac{\partial_\mu f}{1+f} = \frac{1}{2}\overline F_\mu =
\frac{1}{2}\overline \partial_\mu \ln(1+f)\label{eq:89}\\
\Omega_\mu &=& \frac{\partial_\mu\omega}{\omega} =
\frac{1}{2}\overline\Omega_\mu =
\frac{1}{2}\overline\partial_\mu\ln|\omega|.\label{eq:90}
\end{eqnarray}

\section{Dispersion relations}
To determine the dispersion relations for given sources it is useful
to distinguish three cases which will be referred to as the local,
translationally invariant and inhomogeneous cases respectively. In the
local case, the sources are proportional to a delta function. In the
translationally invariant case $A(x,x')=A(x-x')$ and in the inhomogeneous
case $A(x,x')=A(\tilde x,\overline x)$.

There are two ways in which one can proceed with the determination of
the dispersion relations. One is to separate real and imaginary parts
and the other is to used complexified momenta. The latter has several
advantages and makes straightforward contact with the classical theory
of normal modes.  It will be used exclusively for determining the
spectral relations.  Separating real and imaginary parts on the other
hand allows one to identify imaginary contributions as a
Boltzmann/Vlasov equation, illustrating nicely the intimate
relationship between transport and dissipation\cite{calzetta2}.

In order to extract information from the equations it is necessary
to undertake an approximation scheme in which only low order
derivatives are kept in $x$. This is equivalent to an adiabatic
(or quasi-static) scheme in which the development of the system
if slow in comparison to fluctuations, so that fast and slow moving
variables separate in an assumed way. In fact this is already
built into the assumed form of the solution for the Green function,
since without such an assumption, there are no grounds for assuming
that $\tilde x$ and $\overline x$ would separate in the prescribed
manner. For most purposes this approximation is quite sound. For
the present, there seems to be no way of eliminating the approximation.

\subsection{Local sources}
In the local case, the equation satisfied by $H(x,x')$ is
\begin{equation}
\lbrack-{\vcenter{\vbox{\hrule height.4pt\hbox{\vrule width.4pt
height8pt\kern8pt\vrule width.4pt}\hrule height.4pt}}}+m^2+\overline A( x)
-\overline{\partial^\mu\gamma_\mu}
-\overline\gamma^\mu\partial_\mu \rbrack=0.
\label{eq:91}
\end{equation}
Note that, since $\tilde B$ is an odd quantity, it does not
appear in the local limit.
Since one is interested in the variables $x-x'$ and $\overline x$, it
is convenient to Taylor expand $x$ around $\overline x$.
Under the Fourier transform this takes the appearance
\begin{equation}
\overline A(x)H(x,x') \rightarrow \lbrack \overline A(\overline x)
+\frac{i}{2}(\overline \partial_\mu \overline A)\frac{\partial}{\partial k_\mu}
+\ldots\rbrack H(k,\overline x)
\label{eq:92}.
\end{equation}
It is useful to define a new quantity by
\begin{equation}
T_\mu = \frac{\frac{\partial f}{\partial k}}{(1+f)}
\label{eq:93}
\end{equation}
(the steepness of the spectral envelope for the Wigner function)
so that
\begin{equation}
\frac{1}{H}\frac{\partial H}{\partial k_\mu} = T^\mu - v_g^\mu/\omega
\label{eq:94}
\end{equation}
where $v_g=\frac{\partial \omega}{\partial k}$ is the group velocity of the
dispersing wave-packets.
In terms of the quantities (\ref{eq:89}) and (\ref{eq:90})
the action of the spacetime derivative operator on $H(x,x')$ gives
\begin{equation}
\stackrel{x}{\partial}_\mu H(x,x')= 2\pi \int \frac{d^{n-1}k}{(2\pi)^{n-1}}
e^{ik(x-x')}\frac{(1+f)}{2|\omega|}\lbrack ik_\mu+F_\mu-\Omega_\mu\rbrack
\label{eq:95}
\end{equation}
and subsequent derivatives are obtained in a straightforward way.

Substituting $H(x,x')$ into the equation of motion (\ref{eq:91}) now
leads to a second order differential equation for the
frequency $\omega(\overline x)$:
\begin{eqnarray}
\omega^2 &-&2(ik^\mu +F^\mu+\frac{1}{2}\overline\gamma^\mu)\Omega_\mu +
\Omega^2
= {\bf k}^2 +m^2 + \overline A(\overline
x)+\frac{i}{2}(\overline\partial_\mu\overline A)(T^\mu-v_g^\mu/\omega)
\nonumber\\
&-&\overline{\partial^\mu\gamma_\mu}-F^2 -2ik^\mu F_\mu
-\partial^\mu F_\mu -ik^\mu\overline\gamma_\mu - \gamma^\mu F_\mu
\label{insert}
\end{eqnarray}
Clearly this equation presents an insurmountable problem for the
purposes of analytic calculation, thus an approximation
must be made, based on the adiabatic evolution of the average
properties of the system. The lowest order case (which will
be sufficient to reveal the features of interest in this paper) is when
$\Omega_\mu$ and derivatives of $F_\mu$ may be neglected.
This corresponds to a near classical transport of particles, with relatively
few of the fluctuations added by the quantum nature of the field. With
this approximation the dispersion relation may be written:
\begin{equation}
k^2+m^2+\overline A(\overline x)
+\frac{i}{2}(\overline\partial_\mu\overline A)(T^\mu-v_g^\mu/\omega)
-\overline{\partial^\mu\gamma_\mu} -F^2 -2ik^\mu F_\mu -i\gamma^\mu k_\mu
-\gamma^\mu F_\mu=0.
\label{eq:96}
\end{equation}
This can be separated into a more appealing form as the
dispersion relation for a damped oscillator array
\begin{equation}
-\omega^2 -i\Gamma\omega +\omega_0^2=R
\label{eq:97}
\end{equation}
where one identifies the natural frequency,
\begin{equation}
\omega_0^2 = {\bf k}^2 +m^2 -F^2
\label{eq:98}
\end{equation}
the decay constant,
\begin{equation}
\Gamma = -\frac{1}{2\omega}(\overline\partial_\mu\overline
A)(T^\mu-v_g^\mu/\omega)
+\frac{2}{\omega}k^\mu(F_\mu-\frac{1}{2}\gamma_\mu)
\label{eq:99}
\end{equation}
and force term
\begin{equation}
R = \gamma^\mu F_\mu+\overline{\partial^\mu\gamma_\mu}-\overline A(\overline
x).
\label{eq:100}
\end{equation}
One notices how the effective mass of the theory is reduced by the
gradient of the Wigner function $F_\mu$, indicating that rapid transport
could lead to a second order phase transition. This might also
lead to anomalous dispersion.

In a true linear oscillator array $R$, $\Gamma$ and $\omega_0$
would all be independent of the frequency $\omega$. In
equation (\ref{eq:99}) only the zeroth component of the last term is
independent
of $\omega$. This indicates that the decay/amplification of certain modes in
time
is oscillator-like, but that the spatial modes are multiplied by a factor of
${\bf k}/\omega$,  the inverse phase velocity, which has a critical
value when $m/{\bf k}$ is a maximum. This signifies the effect which
a gap in the frequency spectrum can have in leading to anomalous dispersion
in the `plasma'. At high frequencies $\Gamma\rightarrow
k^0(F_0-\frac{1}{2}\gamma_0)$
and the system is oscillator-like. At low frequencies, damping is dominated
by the external potential $\overline A$ and by transport as one might
expect.

\subsection{Translationally-invariant sources}

In the translationally invariant case, all variables are a function of
$x-x'$. One may therefore fully Fourier transform the sources:
\begin{eqnarray}
\overline A(x-x') = \int \frac{d^{n}k}{(2\pi)^{n}}e^{ik_\mu(x-x')^\mu}
\overline A(k)\nonumber\\
\overline \gamma^\mu(x-x') = \int \frac{d^{n}k}{(2\pi)^{n}}e^{ik_\mu(x-x')^\mu}
\overline \gamma^\mu(k)\nonumber\\
\tilde B(x-x') = \int \frac{d^{n}k}{(2\pi)^{n}}e^{ik_\mu(x-x')^\mu} i\tilde
B(k)
\label{eq:101}.
\end{eqnarray}
Note that, since $\tilde B$ is an antisymmetric function a factor of $i$
is introduced to keep $\tilde B(k)$ real.
The equation of motion for $H(x,x')$ is now
\begin{equation}
\lbrack-\stackrel{x}{{\vcenter{\vbox{\hrule height.4pt\hbox{\vrule width.4pt
height8pt\kern8pt\vrule width.4pt}\hrule height.4pt}}}}+m^2 \rbrack H(x,x') +
\int dV_{x''}
\lbrack\overline A(x-x'') -\tilde B(x-x')
-\overline\gamma^\mu(x-x')\stackrel{x''}{\partial}_\mu
- (\overline{\partial^\mu\gamma_\mu})\rbrack H(x'',x')=0.
\label{eq:102}
\end{equation}
The translational invariance enables the latter spacetime integral to be
performed immediately, yielding the dispersion relation
\begin{equation}
k^2+m^2+\overline A(k) -2ik^\mu\gamma_\mu -i\tilde B(k) = 0.
\label{eq:103}
\end{equation}
An apparent consequence of the translational invariance is that $F_\mu=0$
owing to the steady state nature of the system. Comparing the dispersion
relation
to equation (\ref{eq:97}), one identifies
\begin{eqnarray}
\Gamma &=& \frac{2}{\omega}k^\mu\gamma_\mu + \tilde B(k)\nonumber\\
\omega_0^2 &=& {\bf k}^2+m^2\nonumber\\
R &=& - \overline A(k)
\label{eq:104}
\end{eqnarray}
Although the translationally invariant theory describes only
steady state disequilibria, it is nevertheless seen that the
field oscillations are concentrated around the usual mass-shell
$\omega_0^2$ with an amplitude driven by the external force
\begin{equation}
\frac{\overline A(k)}{\lbrack(\omega^2-\omega_0^2)^2
+(\Gamma\omega)^2\rbrack^{\frac{1}{2}}}
\label{eq:105}
\end{equation}
and a quality factor $Q=\omega_0/\Gamma$. Such a steady-state description
would be appropriate for an `infinite laser' i.e. a device which is not
affected by any finite size considerations.

\subsection{Inhomogeneous sources}

The main case of interest is when the sources and Green functions have a
residual dependence on the average position and time. This includes
the local limit as a special case:
\begin{eqnarray}
\overline A(x,x') &=& \alpha(x-x')\beta(x+x')\nonumber\\
\alpha &\rightarrow& \delta(x-x')\nonumber\\
\beta &\rightarrow& \overline A(x) = \overline A(x').
\label{eq:106}
\end{eqnarray}
As usual, one is looking for the eigenspectrum of the quadratic
operator acting on $H(x,x')$. The equation satisfied
by $H(x,x')$ is now:
\begin{equation}
\lbrack-\stackrel{x}{{\vcenter{\vbox{\hrule height.4pt\hbox{\vrule width.4pt
height8pt\kern8pt\vrule width.4pt}\hrule height.4pt}}}}+m^2 \rbrack H(x,x')
+\int dV_{x''}\left[ \overline A(x,x'')-\tilde B(x,x'')-\overline
\gamma^\mu(x,x'')
\stackrel{x''}{\partial_\mu} - \overline{\partial^\mu\gamma_\mu}\right]
H(x'',x')=0
\label{eq:new}
\end{equation}

In the inhomogeneous case there is no dispersion relation
consisting of continuous frequencies in general so the dispersion
relation will only exist for a discrete set. It is convenient to
divide the discussion into two parts: the determination of the
dispersion relation and the nature of the restricted set of values
which satisfy the dispersion relation.

The problem to be addressed is contained in
following form in momentum space:
\begin{equation}
(-{\vcenter{\vbox{\hrule height.4pt\hbox{\vrule width.4pt
height8pt\kern8pt\vrule width.4pt}\hrule height.4pt}}}+m^2)H(x,x')+\int
dV_{x''} \frac{d^nk}{(2\pi)^n}\frac{d^np}{(2\pi)^n}
e^{ik(x-x'')+ip(x''-x')}
S(p,x''+x')H(k,x+x') = \lambda H(x,x').
\label{eq:107}
\end{equation}
The integral over $x''$ is no longer a known quantity in general, but
it is possible to extract an overall Fourier transform
by shifting the momentum $p\rightarrow p+k$ and defining
the average variable of interest $\overline x=\frac{1}{2}(x+x')$:
\begin{eqnarray}
(k^2 &+&
ik^\mu\overline\partial_\mu-\frac{1}{4}\overline{{\vcenter{\vbox{\hrule
height.4pt\hbox{\vrule width.4pt height8pt\kern8pt\vrule width.4pt}\hrule
height.4pt}}}} +m^2)H(k,\overline x)+\nonumber\\
&+&\int dV_{x''}\frac{d^np}{(2\pi)^n} S(k,x+x'')H(k+p,x''+x') = \lambda
H(k,\overline x)
\label{eq:108}.
\end{eqnarray}
In order to find eigenvalues, it is necessary to extract the
factor of $H(k,\overline x)$ from this expression.
This is not possible for arbitrary values of $k$. It is possible,
however,
if the momenta are restricted to a denumerable set expressed by
the property
\begin{equation}
H(k+p,x''+x') = H(k,x''+x')
\label{eq:109}
\end{equation}
which implies that $H(k,\overline x)$ is a periodic function of the
momenta.
The absence of eigenvalues or the failure of this property
leads to the consideration of an infinite iterative mapping
of states, which---in the absence of a stable limit---suggests chaotic
excitations of
the field. This can also be argued geometrically (see the final
section).
Given this mitigating condition, one has
\begin{equation}
\int dV_{x''}\frac{d^np}{(2\pi)^n} S(k,x+x'')H(k+p,x''+x')
= S(k,\overline x)H(k,\overline x).
\label{eq:110}
\end{equation}
The dispersion relation is now obtained in a straightforward
fashion, adopting the same adiabatic approximation
as before, and is given by the implicit relation
\begin{eqnarray}
k^2&+&m^2+\overline A(k,\overline x) +\frac{i}{2}(\overline \partial_\mu
\overline A)(T^\mu-v_g^\mu/\omega)
-i\tilde{B}(k)\nonumber\\
 &-& \overline{\partial^\mu\gamma_\mu}(k,\overline x)
-(F-N)^2 -2ik^\mu(F-N)_\mu -2i\overline \gamma^\mu k_\mu -\overline \gamma^\mu
(F-N)_\mu=0.
\label{eq:111}
\end{eqnarray}
where it is noted that $\lbrace k\rbrace$ is now discontinuous.
Note that the antisymmetry of $\tilde B(k)$ makes it independent of $\overline
x$.
Comparing to the oscillator equation, one has
\begin{eqnarray}
\Gamma &=& -\frac{1}{2\omega}(\overline \partial_\mu\overline
A)(T^\mu-v_g^\mu/\omega)
+\frac{2}{\omega}k^\mu(F_\mu-N_\mu+\overline \gamma_\mu)+\tilde B(k)\nonumber\\
\omega_0^2 &=& {\bf k}^2 +m^2 -(F-N)^2\nonumber\\
R &=& \overline \gamma^\mu F_\mu +
\overline{\partial^\mu\gamma_\mu}(k,\overline x) -\overline A(k,\overline x)
\label{eq:112}
\end{eqnarray}
where $N_\mu$ will be defined presently.

We now turn to the consequences of the condition in equation (\ref{eq:109}).
There are various precedents for such a relation: one is Green functions
defined on a torus (finite temperature, Matsubara formalism,
electron band structure); another is the case of Landau levels on a torus.
The periodicity is clearly the important factor here. In most
of these cases the periodicity is one in real space and the result
is a discrete spectrum of eigenvalues.  Here the periodicity lies in
the momentum itself. In fact the two notions are closely related and
a periodic system in real space has Green functions which are periodic
in momentum space owing to an infinite summation over discrete frequencies
(which is therefore invariant under shifts by a whole number of
periods). The relation (\ref{eq:109}) must be satisfied for all legal
values of the momentum, thus the implication is that the system is
degenerate---i.e. there exist bands of energy which leave the
Green function invariant under certain shifts. These need not all
refer to the same band.
It is therefore possible to write
\begin{equation}
H(k) = H\left(\sum_l \frac{2\pi l^\mu k_\mu}{P_\mu}\right)
\label{eq:113}
\end{equation}
where $P_\mu$ is the {\em momentum} periodicity length (which has dimensions
of inverse space-length).
This finite length must diverge to infinity when the inhomogeneities
vanish. There is only one natural momentum/length scale which has these
properties, namely
\begin{equation}
L_\mu = P_\mu^{-1} = \overline\partial_\mu H(k,\overline x).
\label{eq:114}
\end{equation}
In deriving (\ref{eq:110}) we have used the fact that
\begin{equation}
\int \frac{d^nk}{(2\pi)^n} \exp(ikx)=\delta(x)
\label{eq:115}
\end{equation}
Since $k$ is now restricted to a discrete set, the correctness
of this relation could now be an issue. It can easily be
verified using the formulae
\begin{eqnarray}
\sum_{k=1}^{n} \sin(kx) &=& \sin\frac{(n+1)}{2}\sin\frac{(nx)}{2}{\rm
cosec}\frac{x}{2}\nonumber\\
\sum_{k=0}^{n} \cos(kx) &=& \sin\frac{(n+1)}{2}\cos\frac{(nx)}{2}{\rm
cosec}\frac{x}{2}
\label{eq:116}
\end{eqnarray}
that an extra finite {\em imaginary} contribution can arise from the
discrete nature of the spectrum, which vanishes in the continuous
limit. It will be assumed that such a contribution can be absorbed by
a redefinition of the sources.

Although one is looking at periodic functions, the solution for $H(x,x')$
need not be sinusoidal. In the case of Landau levels on the torus
\cite{laughlin1,burgess6} periodicity is only achieved at the expense of
a flux-quantization condition which, again, involves a degeneracy of
solutions. There is, in fact, an analogy to this situation here.
The extraordinary properties of Landau levels on a torus can be attributed
to the non-translational invariance of the electro-magnetic vector potential.
The similarity here is the non-translational invariance of the many-particle
state as expressed by the dependence on $\overline x$. This point will be
discussed at greater length in the final section, to avoid its meaning
being lost in the present analysis.

The extra terms containing $N_\mu$ can now be explained. They arise from the
$\overline x$ dependence of the momentum space measure:
\begin{equation}
\int \frac{d^nk}{(2\pi)^n} \rightarrow \prod_\mu
\left(\frac{1}{L_\mu}\sum_{l_\mu} \right)
\label{eq:117}
\end{equation}
giving a contribution
\begin{equation}
N_\mu =  \frac{\partial_\mu(L_0\ldots L_{n-1})}{(L_0\ldots L_{n-1})}
\label{eq:118}
\end{equation}
which compounds the non-linearity.
The above restrictions have no special consequences for the Feynman
propagator, since the nature of the momentum is not used to obtain it.
This is gratifying since the Feynman propagator must always be the
literal inverse of the quadratic part of the $+$ time-ordered action. Only the
nature of the singularity is altered in accord with the modified
dispersion:
\begin{equation}
G_F(k) = \frac{1}{-k_0^2+\omega^2-i\epsilon} + 2\pi i f(k,\overline
x)\delta(-k_0^2+\omega^2)
\theta(k_0).
\label{eq:119}
\end{equation}

The appearance of a natural length scale, connected to the inhomogeneities
of a non-equilibrium system, is an important feature for two reasons.
Firstly, the spontaneous generation of a length scale implies the
possibility of domain formation, or a cellular localization in the field.
Secondly, the dependence of the Green functions on themselves implies that
the stable solutions of the system can be regarded as fixed points of
an iterative map. Such maps have been studied in connection with
classical chaotic systems\cite{lauwerier1}.

In the present case, the function $H(k,\overline x)$ depends not
merely on itself but on its derivative. For exponential-like solutions
one could expect that this would amount to the same thing, up to
a constant multiplier. The situation would then be something
akin to $H=H(\lambda H)$, for some constant $\lambda$. This bears a
noteworthy similarity to Feigenbaum's functional equation which
can be written
\begin{equation}
g(x) = \alpha g(g(x/\alpha))
\label{eq:120}
\end{equation}
subject to a boundary value, or rewriting:
\begin{equation}
g(g(\lambda x)) = \lambda g(x).
\label{eq:121}
\end{equation}
This equation has an analytic solution as a power series
\begin{equation}
g(z) = 1 + c_1z^2 + c_2 z^4 + \ldots
\label{eq:122}
\end{equation}
where a limiting value is approached through a geometric
progression with Feigenbaum ration ${\cal F}=4.66$ and
universal scaling factor $\alpha=-2.5$.
Solutions to this equation which fall outside the
fixed point behaviour can be expected to lead to chaotic
behaviour. This strongly suggests that the non-equilibrium
Green functions must exhibit universal behaviour or chaos
in their approach to stable behaviour. In other words, the
approach to equilibrium need not be of the simple damped
or over-damped form of a linear oscillator array---it could
easily entail a chaotic attractor.

\section{Entropy, temperature and the KMS condition}

For systems close enough to a thermal state, it is possible to
define an approximate temperature and entropy. The entropy of the system may
be defined in various ways, often based only on combinatorial
considerations of the micro-canonical picture. Here it is
convenient to define an `oscillator effective entropy' which
is easily related to quantities which arise in the
analysis.
Suppose the
Wigner function is given by the approximate equilibrium form
\begin{equation}
f(k_0,\overline x) = \left( \exp(\beta(\overline x)\omega(\overline
x))-1\right)^{-1}
\label{eq:123}
\end{equation}
then one has
\begin{equation}
\overline F_\mu = - f^2 \left[ \left(\frac{\overline\partial_\mu\beta}{\beta}
+\overline\Omega_\mu\right)\beta\omega\right]
\label{eq:124}
\end{equation}
and, classically, the statistical entropy $S$ is
\begin{equation}
S = k \left( \ln Z + \beta\langle\omega \rangle\right).
\label{eq:125}
\end{equation}
For a harmonic oscillator, one has (see for example ref. \cite{reif1})
\begin{equation}
\ln Z = - {\rm Tr} (1-e^{-\beta\omega}) - \frac{1}{2}\beta\langle\omega\rangle
\label{eq:126},
\end{equation}
thus the oscillator entropy may be defined as
\begin{equation}
S = \frac{1}{2}\beta \langle\omega\rangle + {\rm Tr}\ln(1+f).
\label{eq:127}
\end{equation}
This motivates the definition of a simple measure of entropy for
the oscillator array, given by
\begin{equation}
S_{E}(\overline x) = \int \frac{d^nk}{(2\pi)^n}
\theta(k_0)\ln(1+f)\delta(-k_0^2+\omega^2).
\label{eq:128}
\end{equation}
The rate of change of this entropy is then
\begin{eqnarray}
\overline\partial_\mu S_{E} &=&\overline\partial_\mu
\int \frac{d^{n-1}k}{(2\pi)^{n-1}}
\frac{\ln(1+f)}{2|\omega|}\Bigg|_{k>0}\nonumber\\
&=& \int \frac{d^{n-1}k}{(2\pi)^{n-1}} \frac{(1+f)}{2|\omega|}\lbrack\overline
F_\mu-\overline\Omega_\mu \rbrack
\label{eq:129}
\end{eqnarray}
This quantity can be compared to (\ref{eq:95}). It shows that the entropy
gradient
can be thought of as a `connection' for the field modes. The generation of
entropy is therefore fundamentally connected with the flow of particle
occupation numbers and the `downgrading' of the frequency spectrum---i.e.
the rate at which energy becomes unavailable to do work.

As mentioned earlier, the effect of a non-trivial density matrix, either at
the initial time or later, is reflected in the off-diagonal sources and Green
functions. If one imagines that the sources $A_{\pm\pm}$ arise from a
coupling to another oscillator system\cite{schwinger2} or that they represent
the self-interaction of the field to order $\phi^4$, then $A_{\pm\pm}$ is
essentially the Green function for the field concerned and one would therefore
expect the KMS condition to hold for the sources at equilibrium---now in the
form
\begin{equation}
\theta(|\omega|)A_{+-}(\omega) = e^{\beta\omega}
\theta(-|\omega|)A_{-+}(\omega).
\label{eq:130}
\end{equation}
This condition does not hold in general, but for an
isoentropic process, in terms of the defined quantities at $\gamma^\mu=0$, one
therefore has
\begin{equation}
\tilde B(\omega) = - e^{\beta\omega} \tilde B(\omega)
\label{eq:131}
\end{equation}
It is verified that
\begin{equation}
\frac{\theta(\omega)A_{+-}(\omega)}{\theta(-\omega)A_{-+}(\omega)} =
e^{\beta|\omega|}
\label{eq:132}
\end{equation}
giving $A_{+-}=\sinh(\frac{1}{2}\beta|\omega|)a(\omega)$ for some $a(\omega)$
or
\begin{equation}
B(\omega) = \frac{1-e^{-\beta|\omega|}}{1+e^{\beta|\omega|}} =
\tanh(\frac{1}{2}\beta|\omega|)
\label{eq:133}
\end{equation}
which agrees with Schwinger's result\cite{schwinger2}.
Note that the initial state $f(\overline x_i)$ and its subsequent
development enter only as boundary conditions to the Green functions
and the Wigner function. The changing form of $f(\overline x)$ is
determined solely by the sources $A_{\pm\pm}$. Thus, if the sources
do not evolve, neither does $f(\overline x)$ and neither does
the implicit density matrix. In the perturbation around free
field theory \cite{calzetta1}, $f(\overline x)$ always represents
the state of the system at the initial time.

In the approach to equilibrium one normally expects that dependence on the
average coordinate $\overline x$ to disappear. This is an expression of what is
often called `loss of memory' of the initial state, since $\overline x$ is
measured
relative to the initial time. An equilibrium state (thermal or otherwise)
is, by its nature, either static or periodic, thus the resulting Wigner
function $f(k_0,\overline x)$ must either be independent of $\overline x$ or
a periodic function of this variable. One of the advantages of the present
formulation is that one sees how the sources are responsible for this
loss of memory. Since the sources drive the system, $f(k_0,\overline x)$
can never become $\overline x$ independent as long as the sources are
$\overline x$ dependent. Thus equilibrium will only be secured by
accounting for the back-reaction of the sources to the behaviour of the field.
Explicit equations of motion for the sources have not been considered here.

An example of a periodic `equilibrium' is the case of Rabi oscillations in
the laser (see ref. \cite{firth1} for a review), in which the source and
the field enter into a pendulum-like flip-flop behaviour. An example
of this will be given in the final section.

The decay of field modes is exponential, per mode and is mediated by the
source $\gamma^\mu(x,x')$ and the gradient of the potential $A(x,x')$.
This does not preclude other behaviour for the Wigner function. For example,
in the simplest case close to equilibrium in which the system is quasi-static
and $\overline A = \tilde B = 0$, with almost no external force (see equation
(\ref{eq:112})),
one has $\gamma_\mu\sim F_\mu$ and thus $\partial^\mu F_\mu +F^2\sim0$ giving
$F_\mu \sim x_\mu^{-1}$--- a `long tail' power law decay which parallels
the decay of harmonic waves in curved spacetime\cite{ching1}.

\section{Calculation of expectation values}

The closed time path formalism codifies the causal relationship
between source and response, for the computation of expectation values
in a general mixed state. Since it is redundant except as a
calculational aid, it's introduction should be justified by an
example. The causality of the method is not affected by the
introduction of the sources $A_{\pm\pm}$, but the dissipative dynamics
are.  Normally a fundamental Gaussian theory can never show
dissipation, but in the present situation one has sources which can
drive the field modes and redistribute energy.

There are two cases of interest.
In a self-interacting theory one might identify $A_{\pm\pm}$ with
the correlation function for the field itself $\lambda G_{\pm\pm}$,
giving rise to  dispersion relation of the approximate form
\begin{equation}
k^2+m^2 + \lambda {\rm Tr} (k^2+m^2)^{-1} = 0.
\label{eq:134}
\end{equation}
This is like the variational method used in ref. \cite{calzetta1}.
Lawrie\cite{lawrie1} takes the view that the sources can effect a
renormalization of a self-interacting theory by choosing them in such
a way as to `minimize' the effect of higher order perturbative
contributions. In either case, the effective `resummation' induced by
the sources makes it possible to see damping of field modes at the
one-loop (Gaussian) level.

Consider the response of the field to the source $J(x)$, in the
presence of $A_{\pm\pm}$.  One is interested in the causal expectation
value of the field at time $\overline t$, given the state of the
system at the initial time. The time dependence, in the present
formalism is now contained entirely within the sources---or
equivalently the dispersion relation.  That the CTP generator leads to
a causal result is easily verified by realizing that the expectation
value of the field is always coupled to the sources by the retarded
$n$ point functions.  For an arbitrary action $S[\phi]$,
\begin{eqnarray}
\langle\phi(x)\rangle &=& -i\frac{\delta}{\delta J_+(x)}\Bigg|_{+=-} \langle
0|0 \rangle_{\pm}\nonumber\\
 &=& \frac{1}{2}\int dV_{x'} \lbrack
2G_{++}(x,x')+G_{+-}(x,x')+G_{-+}(x,x')\rbrack J(x') +\ldots\nonumber\\
&=& \frac{1}{2}\int dV_{x'} \lbrack 2G_{++}(x,x')+G^{(+)}-G^{(-)}\rbrack J
+\ldots\nonumber\\
&=& \int dV_{x'} \lbrack G_{++}(x,x')-G^{(-)}\rbrack J(x') +\ldots\nonumber\\
&=& \int dV_{x'} G_{ret}(x,x') J(x')+\ldots
\label{eq:135}
\end{eqnarray}
thus the expectation value depends only on retarded times.
Furthermore, the result is real (being a probability) since the retarded Green
function
is explicitly the real part of the Wightman functions, restricted
to retarded times by a step function:
\begin{equation}
G_{ret}(x,x') = - \theta(t-t') \lbrack G^{(+)}(x,x') +G^{(+)*}(x,x')  \rbrack.
\label{eq:136}
\end{equation}
Making use of the integral representation (\ref{eq:55}), one has
\begin{equation}
G_{ret}(x,x')= -i \int \frac{d\omega}{(2\pi)}\frac{d^nk}{(2\pi)^n}
\frac{\exp(-i\omega(t-t')+i{\bf k}({\bf x}-{\bf x'}))}{(\omega+i\epsilon)}
\left( G^{(+)}(k)+G^{(-)}(k)\right).
\label{eq:137}
\end{equation}
Relabelling and inserting the momentum-space forms for the
Wightman functions from (\ref{eq:69}), one has
\begin{equation}
G_{ret}(k)= - \int d\omega \frac{1}{k_0-\omega+i\epsilon}
\left( \frac{1}{\omega_+}\delta(\omega-\omega_+)-
\frac{1}{\omega_-}\delta(\omega+\omega_-)\right)
\label{eq:138}
\end{equation}
where $\omega_+$ and $-\omega_-$ are the positive and negative frequency
solutions
to the appropriate dispersion relation. These are complex numbers in general,
owing
to the non-vanishing imaginary part labelled as $\Gamma$. Now, since
unitarity demands that $G^{(+)}(x,x')$ be the complex conjugate of
$G^{(-)}(x,x')$,
it is clear that
\begin{equation}
\omega_+^* = \omega_-.
\label{eq:139}
\end{equation}
It is assumed here that the dispersion relation has two complex roots.
The quantity appearing in the delta function in equation (\ref{eq:87})
is then $-k_0^2+\omega_+\omega_-$ which may also be written
$-k_0^2+\omega^*\omega$.
To avoid confusion with previous notation for the absolute value, the complex
modulus will not be denoted $|\omega|$. This indicates that, in spite of the
complex momenta in the dispersion relation, whose role it is to capture
dissipation and transport/kinetic effects, the
`mass shell' constraint is real. The simplest expression
for the retarded Green function is therefore
\begin{equation}
G_{ret}(k)= - \left( \frac{1}{2\omega_+(k_0-\omega_++i\epsilon)}
- \frac{1}{2\omega_-(k_0-\omega_-+i\epsilon)}\right).
\label{eq:140}
\end{equation}
This expression is not manifestly real, since it is a momentum space
result. However, if one defines $2i\tilde\omega = \omega_--\omega_+$ and
$2\overline\omega=\omega_++\omega_-$, where $\tilde \omega$ and
$\overline\omega$ are real, then it is possible to write
\begin{equation}
G_{ret}(k)= \frac{1}{\omega^*\omega}\left(
\frac{(i\tilde\omega k_0 -\omega^*\omega
+2\overline\omega^2)(-k_0^2+\omega^*\omega - 4ik_0\tilde\omega)}
{(-k_0^2+\omega^*\omega)^2+16k_0^2\tilde\omega^2}
\right).
\label{eq:141}
\end{equation}
This may be compared to equation (\ref{eq:105}) and reduces to
\begin{equation}
\frac{1}{-k_0^2+\omega^2}
\end{equation}
when $\omega^* = \omega$ and $\epsilon\rightarrow 0$.  Since the
imaginary part of (\ref{eq:141}) is odd with respect to the momentum
variable $k$, the Fourier transform back to configuration space is
real, as expected.  The desired expectation value is therefore
manifestly real and causal, and the time dependence since the initial
time is contained entirely in the $\overline x$ dependence of the
frequency $\omega$.

\section{Reformulation}

In the preceding sections, it has been shown how dissipation and
amplification of spectral modes can be incorporated into the dispersion
of a quadratic theory, for suitably adiabatic processes. It is
now practical interest to show that the same results can be presented in
another significant form by introducing a `covariant derivative' $D_\mu$
which commutes with the average development of the field state.
This description parallels the structure of a gauge theory
(in momentum space) with an imaginary charge. Alternatively one may
speak of a generalized chemical potential for the `gauge' field.

Consider the derivative
\begin{equation}
D_\mu = \partial_\mu - a_\mu
\label{eq:200}
\end{equation}
and its square
\begin{equation}
D^2 = {\vcenter{\vbox{\hrule height.4pt\hbox{\vrule width.4pt
height8pt\kern8pt\vrule width.4pt}\hrule height.4pt}}} - \partial^\mu a_\mu -
2a^\mu\partial_\mu + a^\mu a_\mu.
\label{eq:201}
\end{equation}
Without any approximation, it is straightforward to show that, in the
general inhomogeneous case,
\begin{eqnarray}
(-{\vcenter{\vbox{\hrule height.4pt\hbox{\vrule width.4pt
height8pt\kern8pt\vrule width.4pt}\hrule height.4pt}}}+m^2)H(x,x') = 2&\pi&
\int \frac{d^{n-1}k}{(2\pi)^{n-1}}\frac{(1+f)}{2|\omega|}
e^{ik(x-x')}\nonumber\\
\lbrace &-& (ik_\mu+F_\mu-\Omega_\mu-N_\mu)^2
-\partial^\mu(ik_\mu + F_\mu -\Omega_\mu-N_\mu)\rbrace.
\label{eq:202}
\end{eqnarray}
It is then natural to make the identification
\begin{eqnarray}
a_\mu &=& F_\mu - \Omega_\mu - N_\mu +\overline \gamma_\mu \nonumber\\
      &=& \partial_\mu S_E(k) - N_\mu +\overline \gamma_\mu
\label{eq:203}
\end{eqnarray}
where the meaning of this notation is such that the expression only defined
when all
objects are under the momentum integration---this is to be understood
in all future expressions.
The field $a_\mu$ is clearly related to the rate of increase of the entropy
$S_E$, the damping factor $\gamma^\mu$ and the rarefaction of the localized
cells $N_\mu$.
One now has:
\begin{eqnarray}
(-D^2+m^2)H(x,x') &=& 2\pi \int
\frac{d^{n-1}k}{(2\pi)^{n-1}}\frac{(1+f)}{2|\omega|}
\lbrace - (ik_\mu -\overline\gamma_\mu)^2) -
\partial^\mu(ik_\mu-\overline\gamma_\mu)\rbrace
\nonumber\\
&=&  2\pi \int \frac{d^{n-1}k}{(2\pi)^{n-1}}\frac{(1+f)}{2|\omega|}
\lbrace k^2 + 2ik^\mu\overline\gamma_\mu -\overline\gamma^2 -i(\partial^\mu
k_\mu)
+(\partial^\mu \overline \gamma_\mu)\rbrace.
\label{eq:204}
\end{eqnarray}
Adding the appropriate source combinations for the inhomogeneous case one has,
without approximation, the differential equation satisfied by $H(x,x')$:
\begin{equation}
\left[ -D^2+m^2 +\overline\gamma^2(k,\overline x) +\overline A(k,\overline x)
- \tilde B(k)+\frac{i}{2}(\overline\partial_\mu\overline
A)(T^\mu-v_g^\mu/\omega) \right]_k H(x,x') = 0
\label{eq:205}
\end{equation}
where the appearence of the subscript $k$ to the
bracket serves to remind that the equation only exists under the
momentum integral. The local limit is simply
\begin{equation}
\lbrack -D^2+m^2-\overline\gamma^2(x) + \overline A(x) \rbrack_k H(x,x')=0.
\label{eq:206}
\end{equation}
The 'gauge' field $a_\mu$ couples via an imaginary unit-charge plays the
role of a generalized chemical potential on the manifold of positive
energy solutions for the real scalar field (the chemical potential has no
meaning
for the full field, since particle numbers are not conserved).
Suppose now that one defines the analogue of the field strength tensor
\begin{equation}
f_{\mu\nu} = \partial_\mu a_\nu - \partial_\nu a_\mu.
\label{eq:207}
\end{equation}
In many cases one will have $f_{\mu\nu}=0$, thus one can 'gauge transform' the
field, which maps
\begin{eqnarray}
\phi(k) &\rightarrow & \phi(k) e^{\int a_\mu dx^\mu}\nonumber\\
        &=& \phi(k) e^{-\int
(F_\mu-\Omega_\mu-N_\mu-\overline\gamma_\mu)dx^\mu}\nonumber\\
        &=& \phi(k) e^{-S_E-\int (N_\mu+\overline\gamma_\mu)dx^\mu}.
\label{eq:208}
\end{eqnarray}
This shows the explicit decay (amplification) of the $k$-th field mode.
The latter relation shows that this process involves an increase
in the effective oscillator entropy of the system.

In terms of the above formulation, the spectral content of the
bosonic theory reduces to the problem of finding the eigenvalues
of the operator $D^2$. In particular one can use the body of experience
gained in the study of gauge theories to attack the problem.
With an adiabatic approximation for $f(\overline x,k)$, $a_\mu$
has a series expansion in powers of $\overline x$. Thus for
quasi static systems
\begin{equation}
a_\mu = (c_0 + c_1 \overline x+\ldots)_\mu.
\label{eq:209}
\end{equation}
The effective field strength
$f_{\mu\nu}$ need not always be zero. Two situations might arise:
(i) the Wigner function might contain a logarithmic singularity, as in
the case where vortices are present, and (ii) the source $\overline\gamma_\mu$
could contain components which specifically drive the macroscopic field
in a given way. A simple example of the latter is the analogue of
Rabi oscillations in the laser, in which the field oscillates between
two states in a regular way. Here, this oscillation is driven by
the source $\overline \gamma_\mu$ or perhaps by a pulsation of the
inhomogeneity scale, and occurs from the linear terms in equation
(\ref{eq:209}).
The current $J=\phi_2 \stackrel{\leftrightarrow}{D}^\gamma \phi_1 $
behaves like a magnetic influence on the system (doing no net work).
Simplifying to the case of a ($1+1$)  dimensional system, one may write
\begin{equation}
a_\mu \sim \overline\gamma_\mu = |\gamma| \epsilon_{\mu\nu}\overline x^\nu
\label{eq:210}
\end{equation}
for constant $|\gamma|$ and $\mu,\nu=0,1$.
This corresponds to a harmonic `flip-flop' motion between field
and source. It is also directly analogous to the well known problem
of Landau levels in an effective magnetic field $|\gamma|$.

The localization in spacetime resulting from the inhomogeneity
scale suggests that such oscillations may take place locally
in cellular regions. A simplified model for this is to impose
periodic boundary conditions on the cells, generating a
kind of global field coherence (this is admittedly
motivated by technical simplicity rather than physical reasoning).
One is therefore led to the study of Landau levels on the
torus---a system which has been studied at some
length\cite{burgess6,laughlin1},
and will not be re-analyzed here.

A significant feature of the Landau problem on the torus is that the
periodicity enforces a flux quantization condition on the field. Here
this translates into the following relation:
\begin{equation}
\overline \partial_0 H(k,\overline x) \overline \partial_1 H(k,\overline x)
| \gamma| = 2\pi n
\end{equation}
for integer $n$. This relation indicates that nearest neighbour
cells might engage in cooperative oscillations (i.e. the size
of cells is quantized in units of the local inhomogeneity scale).
This is clearly a far less stringent condition here than in the
case of a true periodic torus, since the inhomogeneity scale
varies in space and time thus the meaning of strict quantization
is lost. However, it indicates that one can expect a tiling
of spacetime by oscillation cells. Since the size of the cells
might be highly irregular, the tiling behaviour is most likely to be
chaotic unless special geometrical boundary conditions
can enforce a regularity on the field. This is an alternative
expression of the behaviour deduced from the Green function
in equation (\ref{eq:109}).

\section{Conclusion}

Schwinger's closed time path action principle has been applied to the
neutral scalar meson, off-equilibrium, in the presence of long-range,
inhomogeneous sources. The method of dispersion relations is used to
find formal expressions for Green functions which reflect the
absorbative and amplifying processes in the normal modes of the field.
In the case of self-interacting theories, the sources can be thought
of as representing $\phi^{2n}$ interactions to one-loop order, effecting
a resummation of the theory. The effect of rapid transport
(large $F_\mu$) is to induce a change in the sign of the mass
squared, indicating a second order phase transition and anomalous
dispersion.

If significant inhomogeneities or long range interactions exist, the
field naturally forms localized cells with (to lowest order) a
periodic relationship to the natural inhomogeneity length/time scale.
This is shown from the viewpoint of the Green functions and by
recourse to an analogy with Landau levels on the torus.
Since the length scale is determined by non-linear considerations
one can expect chaotic behaviour with islands of order (stable
fixed points) along the approach to equilibrium.
A simple analogue of Rabi oscillations in the laser is shown to
arise as a leading order behaviour in $\overline x$.

The method used in the this work has the advantage of combining the
fundamental aspects of an operator field theory with the usefulness of
the action principle. The use of generating functional ultimately
leads to functional integral forms, as used almost exclusively in the
literature. However, the introduction of the functional integral is
scarcely necessary using the present method and often has the
undesirable effect of turning the discussion of causality into one of
complicated paths of integration in the complex plane.

Comparing to other works reveals both differences and
similarities. Lawrie\cite{lawrie1}, for example, treats the quantity
$\gamma^\mu$ as an explicitly written imaginary part of the spectrum
of excitations. He ignores $F_\mu$, but does not ignore
$\Omega_\mu$. This is a somewhat different approximation which has a
more distant relationship to classical transport theory.  In fact,
since the appearance of $F_\mu$ and $\Omega_\mu$ in $a_\mu$ is
identical, up to a sign, the form of dynamics might well be
independent of the approximation used in this work---understandable as
a reparameterization of an equivalent problem.  Lawrie further
considers $\phi^4$ theory and uses a renormalization-like philosophy
to determine the sources self-consistently thereby effecting a
resummation as noted in equation (\ref{eq:134}).  Calzetta and
Hu\cite{calzetta1} use a variational principle to determine the
effective action for a self-interacting boson theory. This is the same
idea as in ref. \cite{lawrie1}, expressed in extremely aesthetic
formalism and containing important insights into the subject; the
solution to their method is, in practice, more difficult to attain
however and thus results are mainly formal. Neither of these works
consider the implications of non-local effects.  Another interesting
approach is the Schr\" odinger quantization approach in
ref. \cite{eboli1}. This makes a contact with the Schwinger action
principle at a more subtle level and, focusing on somewhat different
issues, uncovers features absent in other formulations of
non-equilibrium physics.

It is important to extend the present analysis to include
both fermions and spin-1 bosons (true gauge fields). The
latter is probably a difficult task in view of the
problems which can arise in gauge fixing. Again, the action principle
approach, starting from the operator field theory is likely
to be the most informative approach.
The appearance of discrete spectra and magnetic like effects
makes the present work very interesting to the study of the
fractional quantum Hall effect. In particular, the pseudo-gauge
field formulation might have interesting connections with the
statistical gauge field employed in the Chern-Simons gauge
theory picture. These and other outstanding issues will be discussed in
future work.

I am grateful to I.D. Lawrie for helpful discussions.

\appendix

\section{Retarded $n$-point functions}

The retarded $n$-point functions are defined by
\begin{eqnarray}
(n=0) ~~~R(x)       &=& = \phi(x) \nonumber\\
(n=1) ~~~~R(x,x_1) &=& = -i \theta(x-x_1) \lbrack\phi(x),\phi(x_1)
\rbrack\nonumber\\
(n>1) ~~~~R(x,x_1\ldots x_n) &=& (-i)^n \sum_{P_i}
\theta(x-x_1)\theta(x_1-x_2)\ldots
\theta(x_{n-1}-x_n) \times\nonumber\\
\left[\left[[\phi(x),\phi(x_1)],\phi(x_2)\right] ,\ldots\phi(x_n)\right]
\end{eqnarray}
where $P_i$ signifies all the permutations of the indices on $x_i$. This
is strictly only defined when all the $x_i$ are different. The coincidence
limit is often defined by recourse to the momentum representation.

The retarded functions have the following properties: (i) $R(x,x_1\ldots x_n)$
vanishes if any $x_i>x$ (with respect to the time), (ii) $R(x,x_1\ldots x_n)$
is a symmetric function of $x_1\ldots x_n$ and (iii) the retarded Green
functions are always defined with respect to a special point $x$ which
is later than all other points.

It may be verified explicitly that the Schwinger-Symanzik generating functional
\begin{equation}
Z_{ret}(x) = \left(T^\dagger e^{-iJ\phi} \right)\frac{\delta}{\delta J(x)}
\left( Te^{iJ\phi}\right)
\end{equation}
generates the $n$-point functions according to the rule
\begin{equation}
R(x,x_1\ldots x_n) = (-i)^n \frac{\delta^n}{\delta J^n} Z_{ret}(x)\Bigg|_{J=0}.
\end{equation}
The step functions are enforced by explicit cancellation of field operators
for times outside
the bounds of the constraints.
The above generating functional is clearly related to the closed time path
generator, and it is easy to see that one may also write
\begin{equation}
R(x,x_1\ldots x_n) \left(\frac{\delta}{\delta J_-}+
\frac{\delta}{\delta J_+} \right)^n (-i)^n \frac{\delta}{\delta J_+}
\langle0|0\rangle_\pm
\Bigg|_{J_{\pm}=0}.
\end{equation}

Finally, it can be observed that that the Hermite polynomials are generated
by the generating functional
\begin{equation}
H_n(z) = (-1)^n e^{z^2} \frac{d^n}{dz^n} e^{-z^2}
\end{equation}
and can therefore be expected to play an important role in the computation of
the transformation function for a quadratic theory.

%\bibliography{noneq}

\begin{thebibliography}{10}

\bibitem{calzetta1}
E.~Calzetta and B.L. Hu.
\newblock {\em Phys. Rev.}, D {\bf 37}:2872, 1988.

\bibitem{pi1}
S-Y. Pi.
\newblock {\em Physica}, A {\bf 158}:366, 1989.

\bibitem{lawrie1}
I.D. Lawrie.
\newblock {\em Phys. Rev.}, D {\bf 40}:3330, 1989.

\bibitem{heinz1}
U.~Heinz.
\newblock {\em Physica}, A {\bf 158}:111, 1989.

\bibitem{korenman1}
V.~Korenman.
\newblock {\em Ann. Phys.}, {\bf 39}:72, 1966.

\bibitem{senitzky1}
I.R. Senitzky.
\newblock {\em Phys. Rev.}, {\bf 119}:670, 1960.

\bibitem{senitzky2}
I.R. Senitzky.
\newblock {\em Phys. Rev.}, {\bf 124}:642, 1961.

\bibitem{brown1}
M.R. Brown and M.J. Duff.
\newblock {\em Phys. Rev.}, D {\bf 11}:2124, 1975.

\bibitem{mottola1}
E.~Mottola in.
\newblock {\em The proceedings from the 3rd workshop on thermal fields and
  their applications}.
\newblock World scientific, Singapore, 1993.

\bibitem{schwinger2}
J.~Schwinger.
\newblock {\em J. Math. Phys.}, {\bf 2}:407, 1961.

\bibitem{eboli1}
O.~Eboli, R.~Jackiw, and S-Y. Pi.
\newblock {\em Phys. Rev.}, D {\bf 37}:3557, 1988.

\bibitem{schwinger1}
J.~Schwinger.
\newblock {\em Phys. Rev.}, {\bf 82}:914, 1951.

\bibitem{bakshi1}
P.M. Bakshi and K.T. Mahanthappa.
\newblock {\em J. Math. Phys.}, {\bf 4}:1, 1963.

\bibitem{calzetta2}
E.~Calzetta, S.~Habib, and B.L. Hu.
\newblock {\em Phys. Rev.}, D {\bf 37}:2901, 1988.

\bibitem{martin1}
P.C. Martin and J.~Schwinger.
\newblock {\em Phys. Rev.}, {\bf 115}:1342, 1959.

\bibitem{burgess99}
M.~Burgess.
\newblock {\em (Unpublished).}, 1994.

\bibitem{laughlin1}
R.B. Laughlin.
\newblock {\em Ann. Phys.}, {\bf 191}:163, 1989.

\bibitem{burgess6}
M.~Burgess, A.~McLachlan, and D.J. Toms.
\newblock {\em Int. J. Mod. Phys.}, A {\bf 8}:2623, 1993.

\bibitem{lauwerier1}
H.A.~Lauwerier in.
\newblock {\em Chaos, edited by A.V. Holden}.
\newblock Manchester University Press, (Manchester), 1986.

\bibitem{reif1}
F.~Reif.
\newblock {\em Fundamentals of statistical mechanics}.
\newblock McGraw-Hill, Singapore, 1965.

\bibitem{firth1}
W.J.~Firth in.
\newblock {\em Chaos, edited by A.V. Holden}.
\newblock Manchester University Press, (Manchester), 1986.

\bibitem{ching1}
E.S.C. Ching, W.M.~Suen P.T.~Leung, and K.~Young.
\newblock {\em Phys. Rev. Lett.}, {\bf 74}:2414, 1995.

\end{thebibliography}

\end{document}